\documentclass{iau}

\usepackage{amsmath}
\usepackage{graphicx}
\usepackage{multirow}
\usepackage{lipsum}

\begin{document}

\lefttitle{K. Vida et al.}
\righttitle{
Detecting coronal mass ejections with machine learning methods
}

\jnlPage{1}{7}
\jnlDoiYr{2021}
\doival{10.1017/xxxxx}

\aopheadtitle{Proceedings IAU Symposium}
\editors{A. V. Getling \&  L. L. Kitchatinov, eds.}

\title{Detecting coronal mass ejections with machine learning methods}

\author{
        K.~Vida$^{1,2}$,
        B.~Seli$^{1,2}$,
        T. Szklenár$^{1,2}$,
        L.~Kriskovics$^{1,2}$,
        A. Görgei$^{1,2,3}$,
Zs.~K\H{o}v\'ari$^{1,2}$ }
\affiliation{{$^{1}$} Konkoly Observatory, HUN-REN Research Centre for Astronomy and Earth Sciences, Budapest, Hungary}
\affiliation{{$^{2}$} CSFK, MTA Centre of Excellence, Budapest, Hungary}
\affiliation{{$^{3}$}  Eötvös University, Department of Astronomy, Budapest, Hungary }

\begin{abstract}
Flares on the Sun are often associated with ejected plasma: these events are known as coronal mass ejections (CMEs). These events, although are studied in detail on the Sun, have only a few dozen known examples on other stars, mainly detected using the Doppler-shifted absorption/emission features in Balmer lines and tedious manual analysis. We present a possibility to find stellar CMEs with the help of high-resolution solar spectra.
\end{abstract}

\begin{keywords}
Sun: coronal mass ejections (CMEs), Stars: activity, methods: data analysis, techniques: spectroscopic, Neural networks
\end{keywords}

\maketitle
\section{Introduction}
On our Sun, most of the larger flares are consistently accompanied by a coronal mass ejection (CME).
CMEs are important because they release massive amounts of plasma into space, impacting space weather and potentially disrupting technological systems on Earth, such as satellites and power grids. Studying CMEs, understanding their dynamics and -- in the future -- developing predictive models are crucial to mitigate potential adverse effects on space-based infrastructure and communication systems.
Solar CMEs are thoroughly investigated through both observational methods and modeling, as outlined in the reviews by 
\cite{2012LRSP....9....3W}  and 
\cite{2017LRSP...14....5K}, among others. The frequency of solar CMEs ranges from 0.5 to 6 CMEs per day, depending on the phase of the solar activity cycle -- these events are observed in detail on the Sun by satellites like the Solar Dynamics Observatory (SDO) or Solar and Heliospheric Observatory (SOHO).

In contrast, the observation of CMEs on other stars poses challenges, and only a handful of such events are known. While flares on these stars can be relatively easily observed through photometry, the observation of CMEs is more challenging. In-situ observations of stellar CMEs will likely remain unfeasible for the foreseeable future.

The primary approach for identifying stellar CMEs involves analyzing their Doppler signature, particularly evident in hydrogen's Balmer lines. The expelled material manifests as a flux increase at the blue wing of the spectral line or, in the case of faster events, may appear as a distinct emission feature, commonly referred to as a "bump" (or absorption if observed against the stellar disk). 
\cite{2017MNRAS.472..876O} extensively discussed the potential observational signatures and constraints.

The majority of reported stellar CME detections in the literature are isolated events discovered serendipitously \citep[e.g.,][]{2016A&A...590A..11V}. 
Until recently, only a limited number of stellar CME detections existed, and the available data were insufficient for robust statistical analysis.
Recent analysis of archive spectral observations has revealed that, although Balmer-line asymmetries are relatively common, 
the occurrence of successful CME events -- those that actually depart from the star -- appears to be less frequent than anticipated based on a straightforward scaled-up solar model \citep{2019A&A...623A..49V,2020MNRAS.493.4570L}.

\section{Using the solar paradigm to understand stellar CMEs}
\begin{figure}[tb]
  \centering
\raisebox{30pt}{  \includegraphics[width=0.25\textwidth]{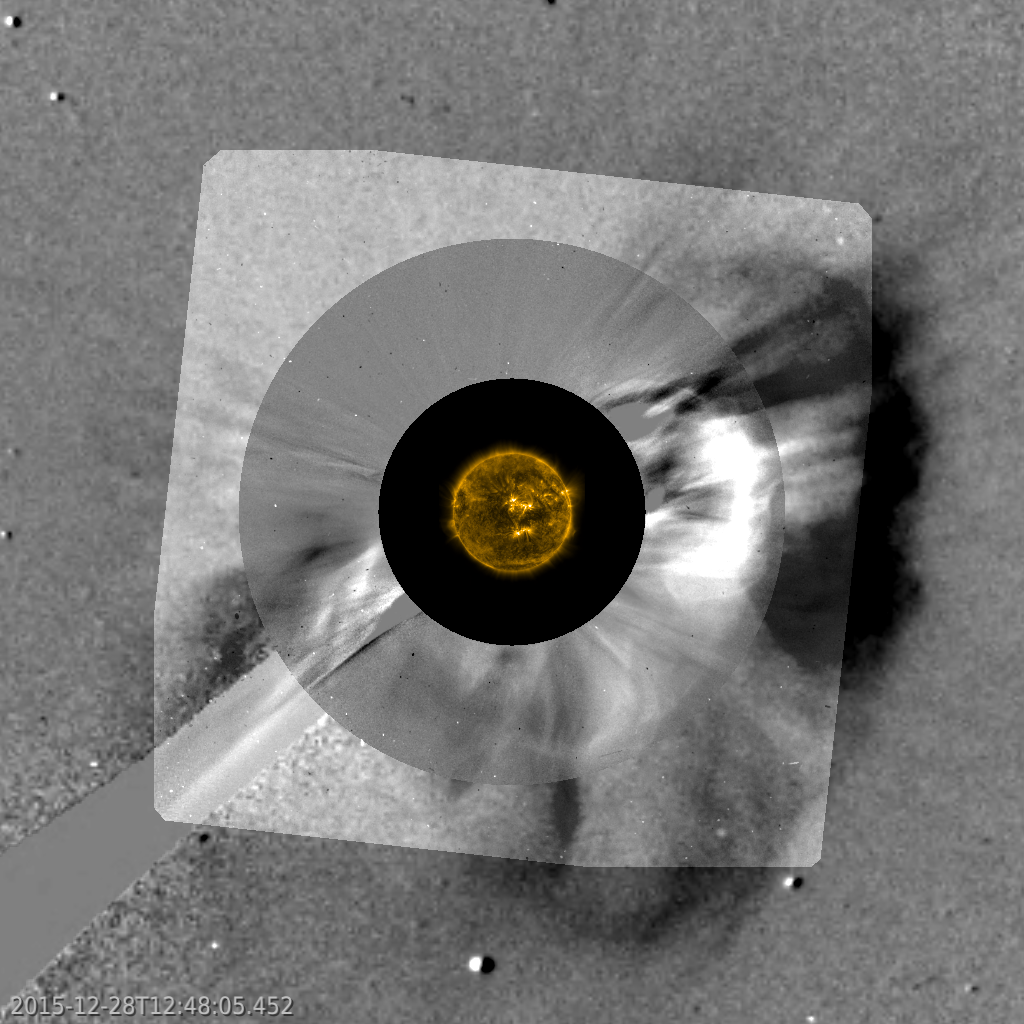} }
  \includegraphics[width=0.70\textwidth]{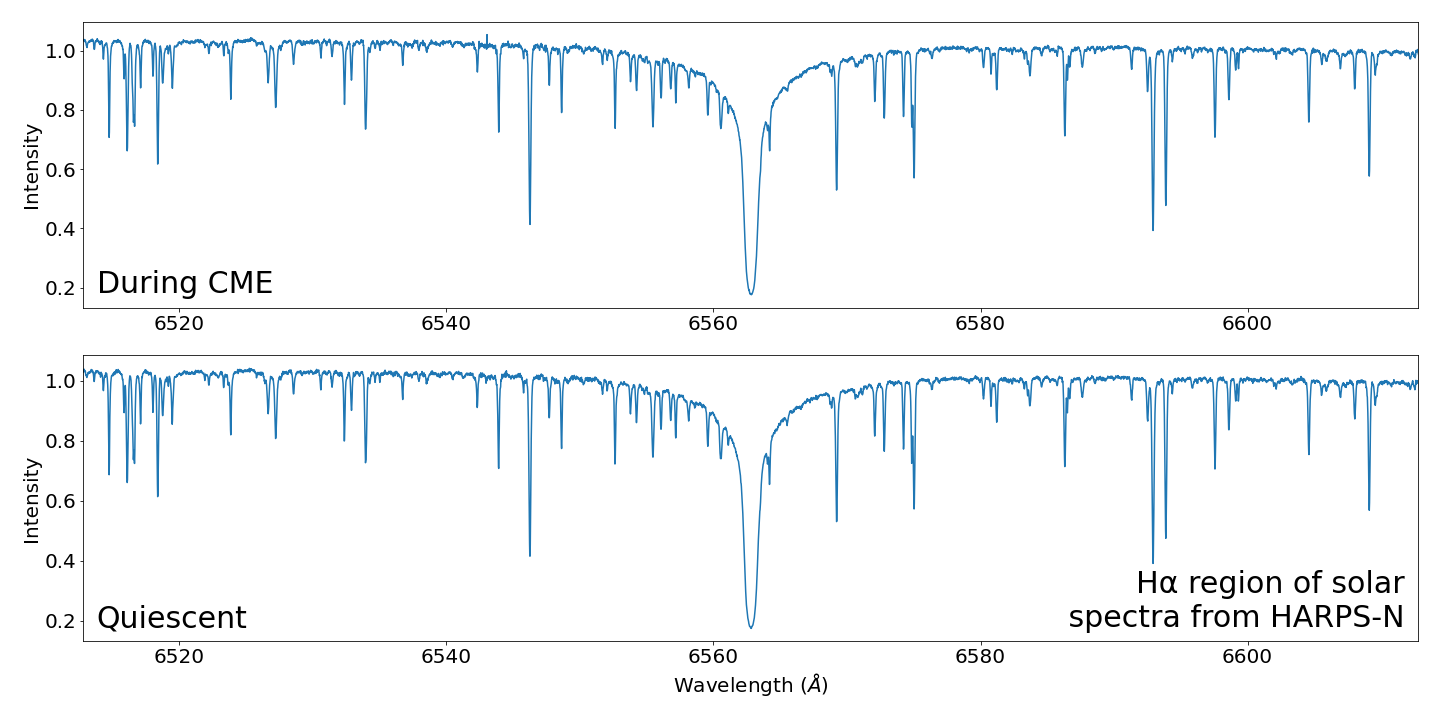}
  \caption{Left: SDO/AIA 171 and SOHO LASCO C2\&LASCO C3 observations of the Sun during a massive CME event. Right: the H$\alpha$ region a solar spectrum during this event compared to the quiescent Sun from HARPS-N observations.}
  \label{fig:dace}
\end{figure}

While a recent investigation of solar-like stars \citep{2020MNRAS.493.4570L} revealed an absence of CME events on these targets, CMEs on solar-like stars do exist \citep{2023arXiv231107380N}. Their rarity could be attributed to either the lower activity levels of these stars, or potential observational biases. 

Explanations for phenomena related to stellar magnetism often draw from the solar paradigm, i.e., assuming that activities on stars are analogous to those observed on the Sun. 
Unlike on stars, on our Sun, we have a detailed knowledge when CMEs occur from direct satellite observations. Could high-resolution, high-cadence Sun-as-a-star observations help us to move forward?
Currently, our Sun is entering a period of increased activity in Cycle 25, anticipated to reach its peak around 2025. During this phase, we expect heightened numbers of flares and increased CME activity, providing a unique opportunity to scrutinize the Sun as a star and refine our understanding of stellar CMEs.

The HARPS-N Solar Telescope is a low-cost solar telescope 
connected to the HARPS-N spectrograph, one of the most precise spectroscopic instruments. The instrument started observing the Sun every clear day with a 5-minute cadence from the Telecopio Nationale Galileo (TNG) facility in La Palma, Spain.
The high-resolution, high-cadence Sun-as-a-star data of the first years of observations are publicly available (see Fig. \ref{fig:dace}).  Using these data with the observations from the STEREO/SOHO solar satellites featuring in-situ CME measurements providing the precise timing and physical properties of CMEs we started developing a neural network to better understand these phenomena. 
If successful, we can use the network to analyze archival spectra possibly revealing events that would not be necessarily found by humans, on other stars.

\section{The first experiments}
\begin{figure}[tb]
  \centering
\includegraphics[width=0.49\textwidth]{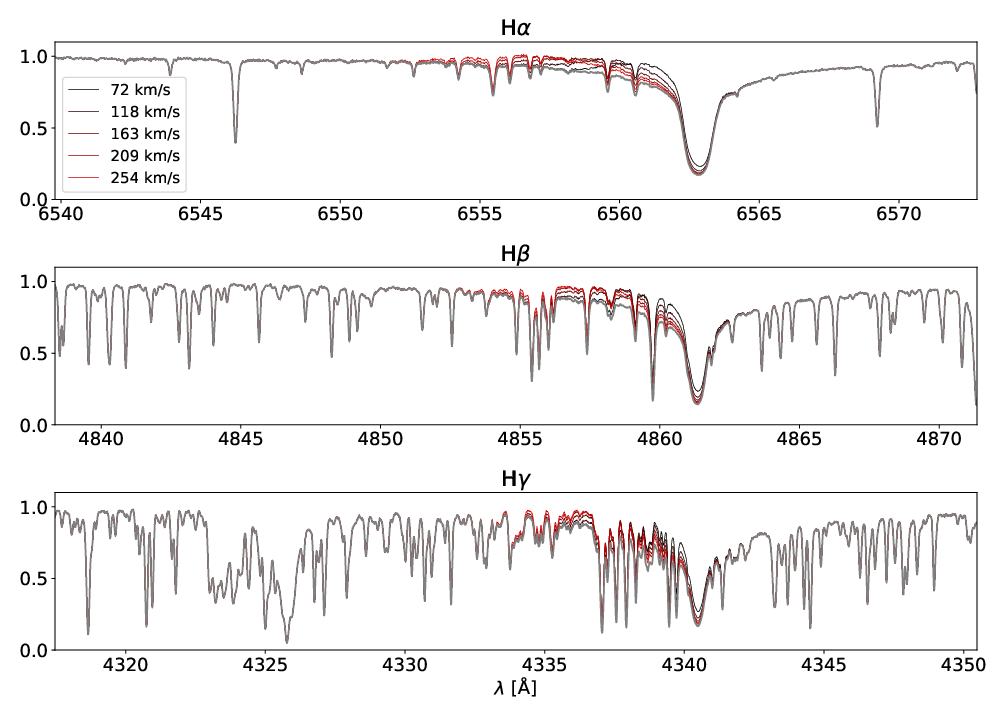} 
\includegraphics[width=0.49\textwidth]{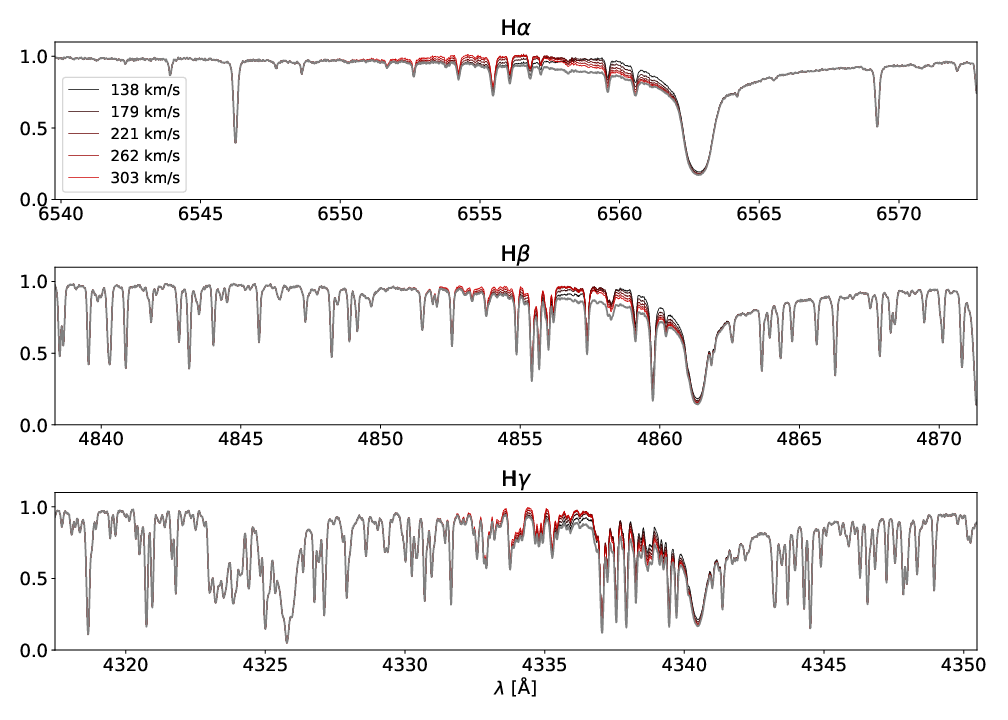}
  \caption{Two examples of injected artificial CME events for training.}
  \label{fig:training}
\end{figure}
First, we checked the optical spectra at the times of the largest solar CME events. Unfortunately, the spectra show no obvious signs of any ejected material as we observe these on other stars (Fig. \ref{fig:dace}) -- the most significant changes in the spectra happen in the telluric lines due to airmass and humidity variations.

In our first experiments, we tried various network architectures using Keras \citep{keras} using both the full spectral range ($\approx$3900--6900\AA) and a smaller region around the H$\alpha$ line as training data, but these returned random predictions and failed to converge.

\begin{figure}[tb]
  \centering
\includegraphics[width=0.45\textwidth]{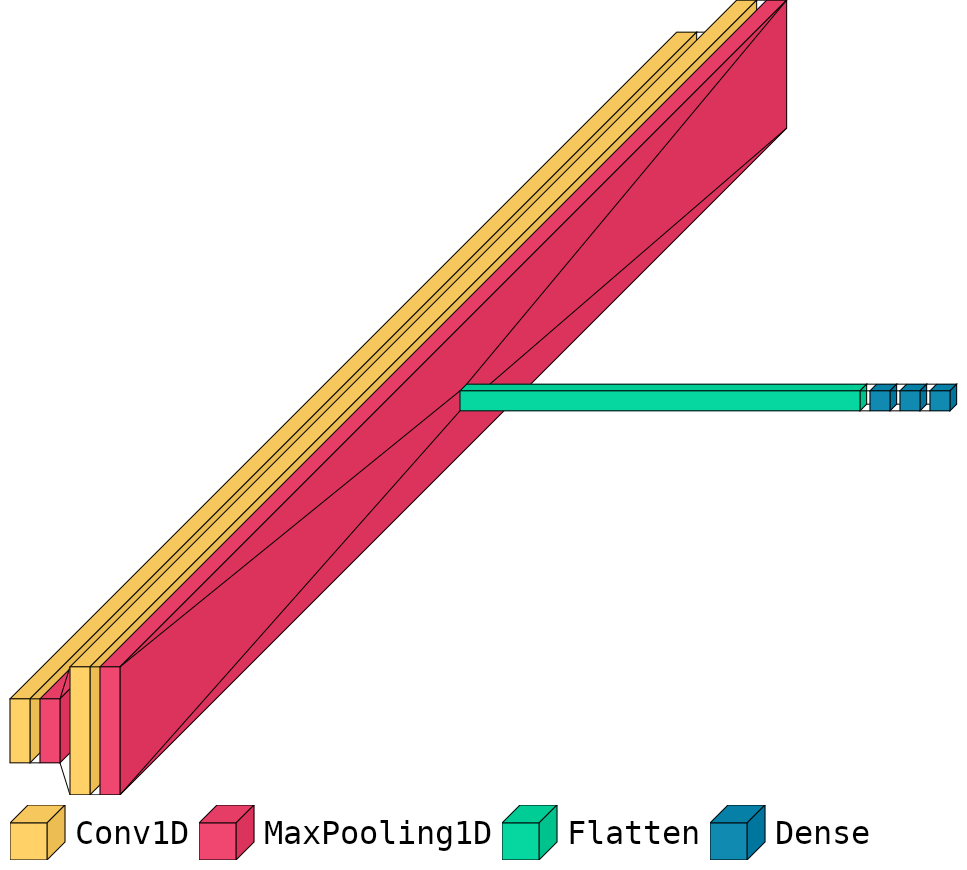} 
  \caption{One of the best-performing network architectures for the artificial CME events.}
  \label{fig:cnn-plot}
\end{figure}

To find out what kind of neural network architecture could be optimal for this problem, we created a training set with artificial coronal mass ejections to have obvious signals. These have a typical mass of $5\times10^{19}$g, 5000 times larger than realistic values. The velocities of the artificial CMEs are realistic, with a maximum velocity of $\approx$400\,km\,s$^{-1}$ (see Fig. \ref{fig:training}). This training set contained 50–50\% spectra with/without CME signals.
{After testing different neural network architectures, the ones based on 1D Convolution performed the best. These types of neural networks "scan" through the data with very high accuracy and great performance.}
After a set of trials, one of the best-performing architecture
-- to our surprise -- was one of the most basic runs, consisting
only two \verb+Conv1D+ layers with 16 and 32 filters, respectively,
followed by two \verb+Dense+ layers with 32 neurons each (see Fig. \ref{fig:cnn-plot}). {During the preparation of the training, validation, and test datasets, we made sure that there was no overlap between these sets.  Our test showed that the accuracy of the deeper 1D Convolution networks was similar but at the expense of speed and performance. Training our neural network was very quick, one epoch took about 2 seconds. For spectral windows of 10\,000 data points (a wavelength range of $\approx$180\AA) the whole training finished in less than 2 minutes.}

While the best runs achieved over 99\% accuracy in both the train and
validation set, this happened only
on artificial data with events 5000 times larger than realistic values
where 50\% of the spectra contained events and
using only 50\,000 data points of the spectrum (the full dataset consists of
more than 200\,000 points).
However, with a network that works consistently on the artificial training
set, we can use transfer learning in the future to use that weight file on the original
scientific observations.
If our experiment is successful, and we can detect coronal mass ejections on
the solar spectra, we can possibly find new CME indicators, and
determine what magnitude of CMEs are observable on other stars. The
new algorithm can be also used to detect CMEs in archive stellar
spectroscopic data.

{Our results based on the artificial data are promising, but of course, there are many other tests ahead of us. Not just the size of the artificial events need to be reduced significantly, but we must utilize the performance of the 1D Convolution to be able to recognize real solar events as well.}

\vspace{1cm}

\noindent{\it Acknowledgements}
Authors from Konkoly Observatory acknowledge the Hungarian National Research, Development and Innovation Office (NKFIH) grants OTKA K-131508 and KKP-143986. LK acknowledges the NKFIH grants OTKA PD-134784. LK and KV are Bolyai János research fellows. KV was supported by the Bolyai+ grant \'UNKP-22-5-ELTE-1093, BS was supported by the \'UNKP-22-3 New National Excellence Program of the Ministry for Culture and Innovation from the source of the National Research, Development and Innovation Fund.

\bibliographystyle{iaulike}
\bibliography{vidaetal_iaus365}

\end{document}